\documentclass[article, onecolumn]{svmult}
\usepackage{geometry}                % See geometry.pdf to learn the layout options. There are lots.
\usepackage{graphicx}
\usepackage{amssymb}
\usepackage{epstopdf}
\usepackage{latexsym}
\usepackage{subfigure}
\usepackage{setspace}
\usepackage{amsmath}
\title{Investigating prostate cancer tumour-stroma interactions - clinical and biological insights from an evolutionary game}
\author{David Basanta\inst{1}\and
Jacob G. Scott\inst{1}\and
Mayer N. Fishman\inst{2}\and
Gustavo E. Ayala\inst{3}\and
Simon W. Hayward\inst{4}\and
Alexander R. A. Anderson\inst{1}}
\institute{
Integrated Mathematical Oncology, H. Lee Moffitt Cancer Center and Research Institute, Tampa, FL 33612, USA 
\and
Departments of Genitourinary Oncology, H. Lee Moffitt Cancer Center and Research Institute, Tampa, FL 33612, USA.
\and
Departments of Pathology, Immunology and Scott Department of Urology. Baylor College of Medicine. 1 Baylor Plz. Houston, TX 77030, USA. 
\and
Departments of Cancer Biology, Urologic Surgery and The Vanderbilt-Ingram Cancer Center. Vanderbilt University Medical Center. Nashville, TN 37232, USA.
\and
Email: david@CancerEvo.org}

\begin{document}
\maketitle
\doublespacing
\section{Abstract}
Tumours are made up of a mixed population of different types of cells that include normal structures as well as ones associated with the malignancy, and there are multiple interactions between the malignant cells and the local microenvironment. These intercellular interactions, modulated by the microenvironment, effect tumour progression and represent a largely under appreciated therapeutic target. We use observations of primary tumor biology from prostate cancer to extrapolate a mathematical model: specifically; it has been observed that in prostate cancer three disparate cellular outcomes predominate: (i) the tumour remains well differentiated and clinically indolent - in this case the local stromal cells may act to restrain the growth of the cancer; (ii) early in its genesis the tumour acquires a highly malignant phenotype, growing rapidly and displacing the original stromal population (often referred to as small cell prostate cancer) - these less common aggressive tumours are relatively independent  of the local microenvironment; and, (iii) the tumour co-opts the local stroma - taking on a classic stromagenic phenotype where interactions with the local microenvironment are critical to the cancer growth. We present an evolutionary game theoretical construct that models the influence of tumour-stroma interactions in driving these outcomes. We consider three characteristic and distinct cellular populations: stromal cells, tumour cells that are self-reliant in terms of microenvironmental factors and tumour cells that depend on the environment for resources but can also co-opt stroma. Using evolutionary game theory we explore a number of different scenarios that elucidate the impact of tumour-stromal interactions on the dynamics of prostate cancer growth and progression and how different treatments in the metastatic setting can affect different types of tumors.  

\section{Introduction}
When detected early, prostate cancer is a largely curable entity.  Once metastatic, like many cancers, cure becomes impossible and one is left with a strategy of chronic disease management.  Metastatic prostate cancer is initially managed by therapeutic manipulations aimed at ablating endogenous androgen production,  a strategy that deprives cells of growth signaling factors.  Additionally, bisphosphonates or RANK ligand inhibitors disrupt tumor-stroma interactions in the bone and are routinely considered as part of the standard of care for metastatic disease \cite{Gallo:2011fk}.  Given a sufficiently long time interval, the emergence of resistance to these strategies is inevitable - for example, with hormone therapy resistance, this constitutes a situation called castrate resistance.  Once resistance emerges, the disease becomes much more difficult to control, symptoms worsen, and the life expectancy drastically shortens.

Carcinogenesis and cancer progression result from evolutionary processes in which the interactions between tumour cells, their environment and the surrounding stroma results in proliferation of cells from the genetically unstable tumor and, consequently, clinically malignant behavior. The concept of microenvironmental changes associated with, and potentially promoting expansion of the clones with phenotypes that define cancer dates back to observations by Galen in the second century \cite{reedy:1975}. Bone marrow derived cells, widely distributed through tissues, play a complex range of roles including generating immune and inflammatory responses and contributing to fibrotic changes. 

The role of specific cell types within this complex hierarchy is becoming better understood. Inflammatory cells have been shown to be involved in tumour promotion at many sites including the prostate \cite{Balkwill:2009if,De-Marzo:2007gd,Mantovani:2008jy}. This is likely due to the stimulation of persistent proliferative conditions in the presence of a mutagenic environment: some associated factors include the local activation of reactive oxygen species \cite{Kundu:2008bf,Schetter:2009qw}. 

Well known examples of stromal cells that change in response to the presence of a tumour include fibroblasts and osteoclasts, among others. In a series of papers in the 1980s, Schor and co-workers demonstrated that fibroblasts adjacent to carcinoma epithelium were fundamentally different from normal stroma and that these changes were implicated in neoplastic progression \cite{Schor87}. These malignancy-associated changes only occurred in a subset of the resident fibroblasts \cite{Schor:1987lt,Schor:1988vz}.  References specifically to tumour-associated or carcinoma-associated fibroblasts, myofibroblasts and reactive stroma become abundant in literature from the 1970s onward. 

The relationship between osteoclasts and tumor cells in metastatic disease in the bone has been of recent interest, and much research has gone into understanding signaling cascades including many of the matrix metalloproteinases\cite{Lynch:2005uq} which seem to promote a vicious cycle of bone turnover and tumor promotion.  Sadly, while these pathways are reasonably well understood, the clinical application of our knowledge in this realm has had little impact with the notable exception of cladronate in breast cancer - a case where a 'stromal directed' therapy - not an anticancer agent - has actually been shown to increase survival\cite{Diel:2008fk}.

\begin{figure}[!htb]
   %\centerline{\includegraphics[width=8cm]{stromagenic.png}\includegraphics[width=8cm]{stroInd.png}}
   \centerline{\includegraphics[width=16cm]{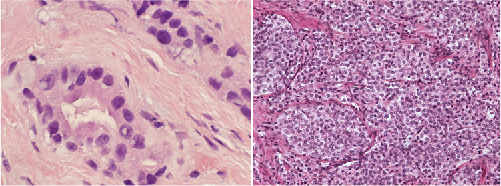}}
\caption{\label{fig:real}Lethal prostate cancer phenotypes. On the left we show a glandular tumour with abundant reactive stroma (stromogenic carcinoma), the image on the right is a poorly differentiated cancer without intervening reactive stroma (stroma independent tumour).}
\end{figure} 

We, and others, have shown that carcinoma associated fibroblasts derived from human prostate tumours can promote tumourigenesis \cite{Barclay:2005ta,Olumi:1999km,Kiskowski:2011uq,Franco:2011fk}. We have also demonstrated that the stromal phenotype in a tumour can be used as a basis for patient disease progress prognostication \cite{Ayala:2003zh,Yanagisawa:2007tp}. While some of the pathways underlying the ability of cancer stroma to regulate tumourigenesis have been elucidated \cite{Ao:2007fx,Ao:2006ud,He:2007yj}, the situation is complex and many interactions remain to be explored \cite{Bierie:2006qz}, especially with regards to the progression of the carcinoma towards either stromagenic (figure \ref{fig:real} left) or stromal independent outcomes (figure \ref{fig:real} right). 

In this paper we introduce an Evolutionary Game Theory (EGT) model that studies the evolution of three different cell populations over time: stromal cells, a dependent tumour phenotype capable of co-opting stromal cells to support its growth, and an independent tumour phenotype that does not require microenvironmental support, be it stromal associated or not. This model is then applied to the clinical problem of metastatic prostate cancer.

\section{The game}
In EGT the behaviour of the players is not assumed to be based on rational payoff maximisation but  it is thought to be shaped by trial and error - adaptation through natural selection or individual learning \cite{Maynard:1982}. In the context of the evolution of populations there are two game theory concepts to be interpreted differently than in traditional game theory. First, a strategy is not a deliberate course of action but a phenotypic trait. The payoff is Darwinian fitness, that is more average reproductive success. Secondly, the players compete or cooperate to become a larger share of the population \cite{Sigmund:1999}. 

Genetic and epigenetic changes can transform the cells in a co-operating healthy tissue ecosystem.  One change is to produce individualistic tumour cells that compete for space and resources \cite{nowell:1976,Crespi:2005,Merlo:2006} and which can attract support from other cell types, for example by replicating developmental scenarios where cell growth is a normal outcome. If we consider a tumour as an ecosystem, it is possible to use tools from ecology, such as EGT, to study the evolution of the different cellular populations.  EGT has been used to explore various aspects of glioma progression \cite{basanta:2008a,Basanta:2011fk},  the influence of the tumour-host interface in colorectal carcinogenesis \cite{gatenby:2003a}, the role of phenotypic variability in multiple myeloma \cite{dingli2009} and the evolution of a number of phenotypic traits such as motility and invasion  \cite{Mansury:2003,Basanta:2007} or microenvironmental independence \cite{Anderson:2009eu}.

As in our previous work \cite{Anderson:2009eu}, the model assumes a tumour with two different epithelial phenotypes: tumour cells that depend on the microenvironment for survival ($D$) and those that are independent of the microenvironment ($I$). Table \ref{tab:payoffs} shows the payoffs for each cell type when interacting with others. A further assumption is that no other phenotypes are relevant in the context of the game and that spatial considerations will not affect the outcome \cite{Hofbauer:1998}. The payoffs in EGT represent the fitness change resulting from the interaction - a positive change represents an increase in the long-term growth rate of the cell. The payoff values are normalised in the range [0:1] so 1 represents the maximum fitness for any given phenotype.

\begin{table}[h]        
	\begin{center}
	\caption{Payoff table that represents the interactions between the three cell types considered in the model. The fitness of each of the phenotypes ($S$: Stroma, $D$: microenvironmentally dependent, $I$:microenvironmentally independent) depends on the interactions with other phenotypes and the values of the costs and benefits resulting from these interactions. These costs and benefits are $\alpha$ (benefit derived from the cooperation between a $S$ cell and a $D$ cell), $\gamma$ (cost of being microenvironmentally independent), $\beta$ (cost of extracting resources from the microenvironment and $\rho$ (benefit derived by $D$ from paracrine growth factors produced by $I$ cells. }
\begin{tabular}{|c|c|c|c|} \hline
	  &      S & D          & I     \\ \hline
	S            &   0 &      $\alpha$ & 0                         \\ \hline       
	D             &   $1+\alpha-\beta$ &  $1-2\beta$ & $1-\beta+\rho$       \\ \hline
	I &$1-\gamma$  & $1-\gamma$ & $1-\gamma$ \\ \hline
	\end{tabular}  
	\label{tab:payoffs}
	\end{center}
\end{table}

The $I$ cells produce their own growth factors (e.g. testosterone) and space (e.g. by degradation of the extra cellular matrix) and thus are considered to have a relatively constant fitness ($1-\gamma$), where $\gamma$ represents the fitness cost for $I$ cells to be independent, instead of committing those resources to faster proliferation.  The $D$ cells rely on their microenvironment for survival and growth at a fitness cost ($\beta$) that represents, among other things, the scarcity of resources and space. A nutrient deprived microenvironment would then be characterised by a higher value of $\beta$. Since $I$ cells produce space and shareable growth factors, this model assumes that $D$ cells derive a fitness advantage from their interactions with $I$ cells represented by the variable $\rho$. On the other hand, $D$ cells interacting with other $D$ cells will have a harder time sharing resources with equally dependent cells and thus are assumed to have double the cost $2\beta$ for relying on the microenvironment for survival and growth and thus have a fitness of $1-2\beta$.

We further consider a stromal population ($S$) that can interact with the tumour. The stromal compartment is thought to be complicit in tumour progression and represents a potential target for new therapies \cite{Strand:2009uq} as well as possibly an underrecognized one for current therapies such as androgen ablation and bisphosphonates.  Stromal cells retain  an ability to undergo rapid proliferation but normally are relatively growth quiescent with low rates of proliferation and death. For this reason the fitness benefit derived by stromal cells from the interactions with tumour cells is assumed to be zero. However, under certain circumstances, stromal cells are susceptible to being co-opted by certain tumour phenotypes (much like carcinoma associated fibroblasts). In this situation, stromal and tumour cells produce factors that stimulate each other's growth in a mutualistic fashion. For the EGT model this is represented by the variable $\alpha$ in the payoff table. A low $\alpha$ represents tumours in which the stroma cannot be co-opted.

If $p_{t}^{I}$ is the proportion of $I$ cells at a given time $t$ and $p_{t}^{D}$ the proportion of $D$ cells then the absolute fitness of each cell population ($W(S)$, $W(I)$, $W(D)$) will be given by the following expressions:

\begin{equation}  W(S) = p_{t}^{D}\alpha, \end{equation}
\begin{equation} W(I) = 1 - \rho, \end{equation}
\begin{eqnarray} W(D) = (1-p_{t}^{I}-p_{t}^{D})(1-\beta+\alpha)+p_{t}^{I}(1-\beta+\rho)+p_{t}^{D}(1-2\beta) \\
   =  1-\beta+\alpha+p_{t}^{I}(\rho-\alpha)-p_{t}^{D}(\beta+\alpha).\end{eqnarray} 

The average fitness ($\bar{W}$) of the population is given by:
\begin{equation}
	\bar{W}=(1-p_{t}^{I}-p_{t}^{D})W(S)+p_{t}^{I}W(I)+p_{t}^{D}W(D)
\end{equation}

From these expressions it is possible to derive the discrete replicator equations that describe, using the absolute fitness of each of the populations, how they change over time \cite{Maynard:1982}. The proportion of a cellular population in the model at a given time $t$ will depend not only on its own fitness (W) but also on the fitness of the other cell populations. If the fitness of a phenotype X, $W(X)$ is higher than the average fitness of all the phenotypes combined ($\bar{W}$) then the proportion of that phenotype will increase during the next time step, for as long as the reasons that keep the phenotype relatively fit remain. The replicator equations are:

\begin{align}
p_{t+1}^{I}=p_{t}^{I}\frac{W(I)}{\bar{W}},\\
p_{t+1}^{D}=p_{t}^{D}\frac{W(D)}{\bar{W}}.
\end{align}

\section{Results}
One may apply the replicator equations described in the previous section to study the temporal evolution of the different populations in a number of scenarios. These scenarios are characterised by the 4 variables of the model: $\alpha$, the mutual benefit between $D$ cells and co-opted stroma, $\rho$, the benefit that $D$ cells derive from coexisting with $I$ cells (which can produce growth factors and space), $\beta$, the fitness cost of relying on a potentially unfriendly microenvironment (e.g. scarce resources or susceptibility to environmental apoptotic signals) and $\gamma$, the cost that $I$ cells have to incur in order to become independent from the microenvironment. 

We considered three scenarios describing nutrient rich, neutral and poor environments (assigned as three values of the fitness cost of dependence, $\beta$ = 0.2, 0.5 and 0.8).  In each case, the replicator equation was iterated 20 times from an initial condition assuming that both tumour populations represented a very small proportion, each, $10^{-4}$ of the population, with the rest being stromal cells ($S$). Different initial conditions yield similar results although the time required to reach equilibrium depends on the proportion of tumour cells in the population. We chose 20 time steps as it appears to be sufficient for most simulations to reach stability.

\begin{figure}[!htb]
   \centerline{\includegraphics[width=20cm]{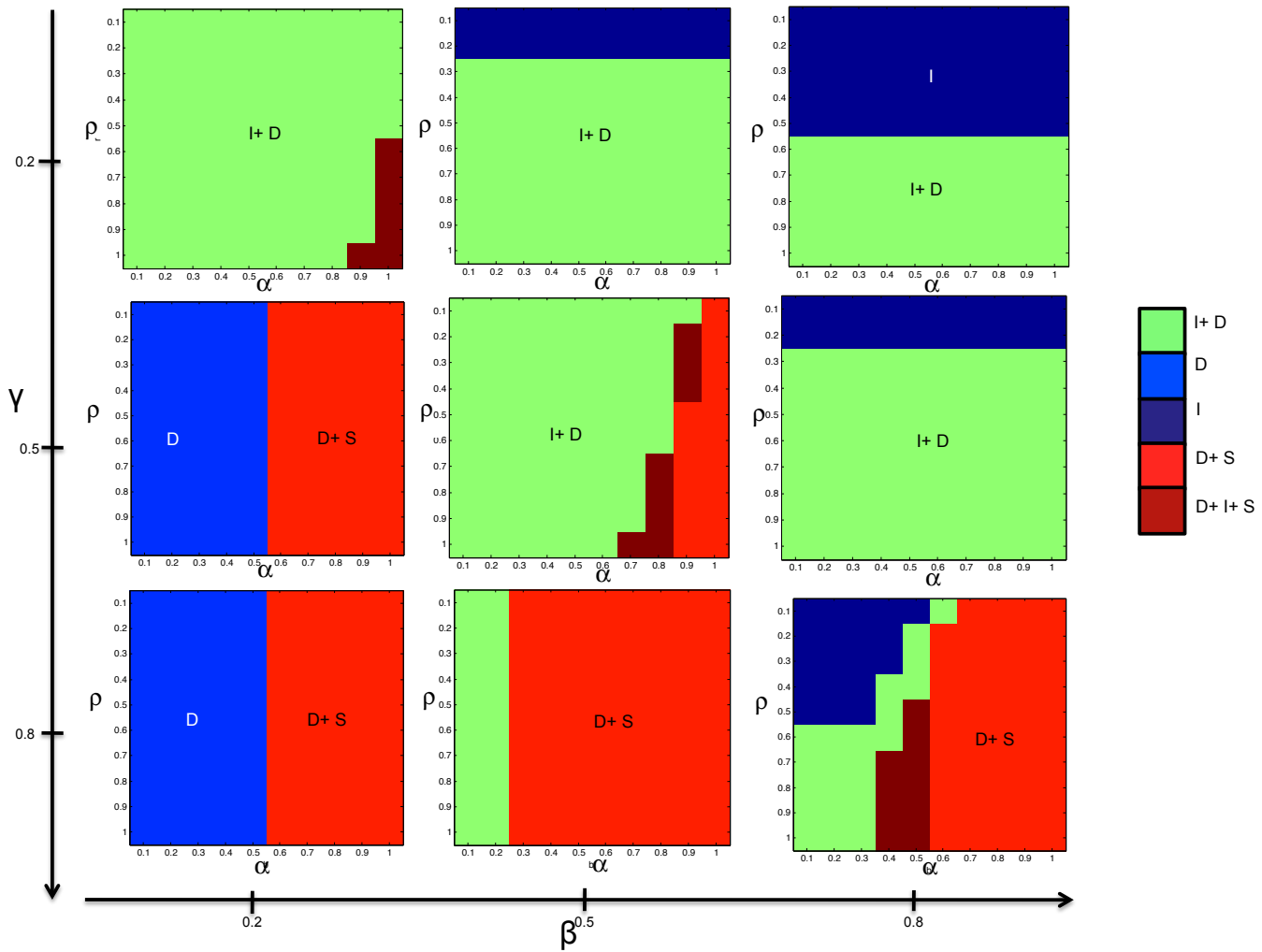}}
\caption{\label{fig:outcomes} Outcomes from the replicator equations under a range of costs associated with $S$, $I$ and $D$ phenotypes. Each box represents the outcome of the replicator equation in which the specific values of $\alpha$ and $\rho$ are varied (from 0.1 to 1). We are assuming three types of scenarios in terms of availability of resources and space: from rich (left column, characterised by $\beta=0.2$) to medium (centre column, $\beta=0.5$) to poor (right column, $\beta=0.8$). We have also hypothethised three different I phenotypes, from very fit (upper row, with $\gamma=0.2$) to medium (centre row, $\gamma=0.5$) and unfit (lower row, $\gamma=0.8$).} 
\end{figure} 

The first row in figure \ref{fig:outcomes} shows the outcomes from three scenarios characterised by different costs of dependence on the microenvironment with a baseline $I$ phenotype cost set to $\gamma=0.2$. When the microenvironment is rich in resources ($\beta = 0.2$), the main outcomes are either coexistence of tumour phenotypes or, when $\alpha$ or $\rho$ is high (increasing the fitness of the $D$ population), coexistence of the three phenotypes.  The second row shows the outcomes from three scenarios characterised by different costs of relying on the microenvironment with an $I$ phenotype with $\gamma=0.5$.  When the microenvironment is rich in resources the main outcomes are either dominance of $D$ phenotypes or, when $\gamma$ is sufficiently high, coexistence of $D$ and $S$. 

When resources are neither scarce nor plentiful the main outcomes are coexistence of $I$ and $D$ if $\alpha$ is sufficiently small or coexistence of $I$ and stroma if $\alpha$ is high enough. Interestingly for some values of $\alpha$ (from medium to intermediate-high), the outcome tends to be coexistence of the three phenotypes. For environments poor in resources ($\beta = 0.8$) the main outcomes are driven by $\rho$, with low values of $\rho$ (benefit for $D$ cells coexisting with $I$ cells) leading to dominance of $I$ phenotypes and higher ones leading to coexistence of both tumour phenotypes. 

%**In the left column; when resources are scarce (high $\beta$) the main outcomes are coexistence of $I$ and $D$ if $\rho$ is sufficiently small or coexistence of $I$ and $S$ if $\rho$ is high.

The third row shows the outcomes from three scenarios characterised by different costs of relying on the microenvironment with an $I$ phenotype with $\gamma=0.8$, representing a relatively unfit $I$ population.  When the microenvironment is rich in resources the main outcomes are either dominance of $D$ phenotypes or, when $\alpha$ is sufficiently high, coexistence of $D$ and $S$. When resources are neither scarce nor plentiful the main outcomes are coexistence of $I$ and $D$ if $\alpha$ is sufficiently small or coexistence of $D$ and $S$ if $\alpha$ is high enough. Perhaps the most dynamically intersting and biologically relevant sumulation is when the microenrionment is poor in resources ($\beta = 0.8$) and the cost of being independent is high ($\gamma = 0.8$) as shown in the bottom right corner of \ref{fig:outcomes}.

If both $\rho$ and $\alpha$ are low then the tumour will be dominated by $I$ phenotypes. If $\alpha$ is  0.5 or below, a sufficiently high value of $\rho$ leads to tumours that contain both $I$ and $D$ cells. When $\alpha$ is higher than 0.5 the cooperation between stroma and $D$ results in the extinction of $I$ cells. Coexistence between the three phenotypes occurs when $\rho$ is high (which helps the $D$ cells) and $\alpha$ is below 0.5 but not too low (over 0.3) which promotes cooperation between $D$ and stromal cells without driving $I$ cells to extinction.

\begin{figure}[!htb]
   \centerline{\includegraphics[height =7cm,width=9cm]{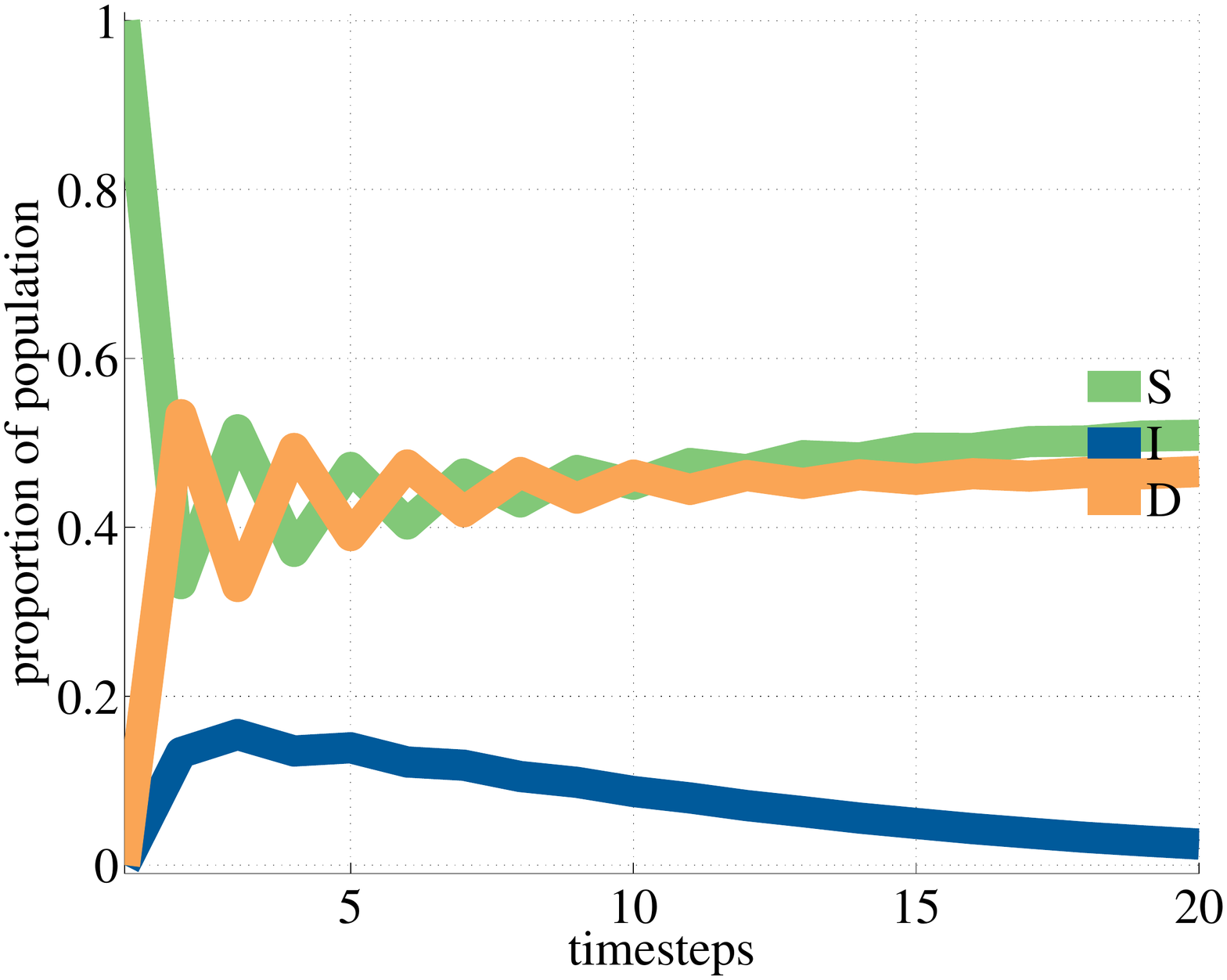}\includegraphics[height =7cm,width=9cm]{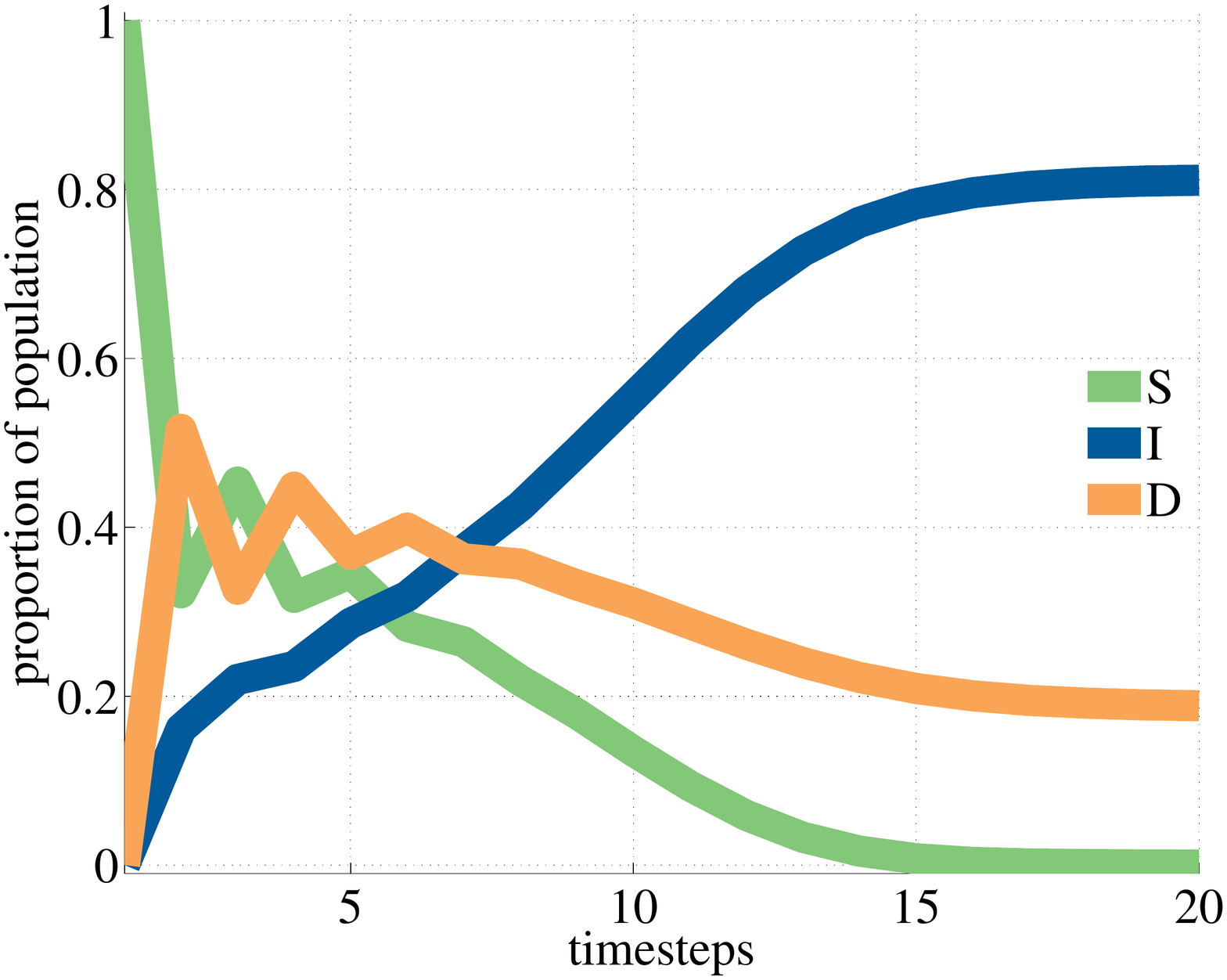}}
\caption{\label{fig:example}\textbf{Left}. Stromagenic tumour with coexistence of $D$ and stromal phenotypes. In this example the stromal population derives a benefit from its interaction with the $D$ cells . As a result the $I$ population is driven to extinction. The initial proportion of $I$ and $D$ cells is  $10^{-4}$ with the rest being stromal cells. The parameters characterising the game were $\alpha=0.5, \beta=0.7, \rho=0.1$ and $\gamma=0.8$. \textbf{Right}. Stromal independent tumour with dominance of $I$ phenotypes. In this example the stromal population derives a benefit from its interaction with the $D$ cells but both populations are still outcompeted by the Is. As a result the stromal population is driven to extinction and the $D$ to irrelevancy. The initial proportion of $I$ and $D$ cells is  $10^{-4}$ with the rest being stromal cells. The parameters characterising the game were $\alpha=0.5, \beta=0.7, \rho=0.1$ and $\gamma=0.75$.}
\end{figure} 

As the emergence of tumours dominated by stroma (stromagenic) versus those independent of stroma is of greatest interest, the dynamics of these outcomes were explored in more detail.  Figure \ref{fig:example} shows the time evolution of the replicator equation for two specific examples. The figure shows how small changes in the fitness of $I$ (as determined by $\gamma$) or $D$ phenotypes (as determined by $\beta$, $\rho$ and $\alpha$) could result in large changes in the population dynamics leading to fundamentally different outcomes. Left panel: a sufficiently high value of $\alpha$ and low value of $\rho$ means that $D$ cells derive a much higher benefit from their cooperation with stromal cells than from their interactions with $I$ cells. Given the low general fitness of the I cells (with $\gamma=0.8$) it is not surprising that they quickly become extinct as the advantage of cooperation sustains and promotes $D$ and $S$ cells in the tumour. The right panel in the same figure shows a slightly fitter $I$ phenotype (with $\gamma=0.75$ instead of 0.8), the $I$ population manages to sustain growth and, as it becomes an increasingly larger part of the tumour population, disrupts the already initiated cooperation between $D$ and $S$ cells, resulting in the extinction of the stromal population and a $D$ population that represents a smaller part of the tumour compared with the previous example.

\begin{figure}[!htb]
   \centerline{\includegraphics[width=12cm]{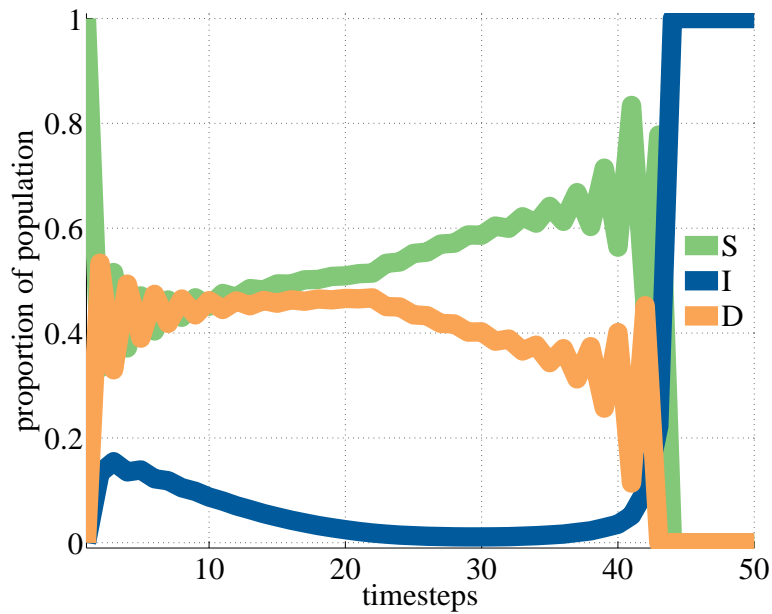}}
\caption{\label{fig:discussion} Microenvironmental perturbation leading to a switch from stromagenic to a stromal independent tumour. Starting with the situation described in figure \ref{fig:example} left, where  $\alpha=0.5, \beta=0.7, \rho=0.1$ and $\gamma=0.8$, after iterating the replicator equation for 20 times, $\beta$ was increased by 0.1 to 0.8 to signify a poorer microenvironment. As a result, the $D$-$S$ cooperation is disrupted and an equilibrium resembling that shown in figure \ref{fig:example} right is reached.}
\end{figure} 

In reality, changes in the microenvironment can occur during tumor progression.  Figure \ref{fig:discussion} illustrates how a dynamic microenvironment can disrupt the outcome and drive the tumour from stromagenic to stromal independent or viceversa.  The replicator equation equations are initially iterated to form a stromagenic tumour (as in the left panel of figure \ref{fig:example}) then after 20 iterations the microenvironment is altered such that it becomes harsher (i.e. increase $\beta$ from 0.8 to 0.9). This results in a destabilisation of the cooperation between the $D$ and $S$ populations and a state transition to one exhibiting dominance of the $I$ subpopulation.

\section{Therapeutic implications}

From the results presented in figure \ref{fig:example} it is clear that small changes in the fitness of the $I$ phenotypes (given all other parameters are equal) can lead to a large scale changes in populations - even causing phase transitions from stromal independence to a stromagenic tumours. While both of these tumor types are eventually lethal, it has been postulated that the stromagenic tumours have a longer natural history\cite{Ayala:2003zh}\cite{Ayala:2011kx}.  Further, because of the biological aspects of the different phenotypes, they are likely sensitive to different types of therapy.  For example, manipulation of a biological pathway (such as mTOR) would differentially penalise cells which depend more upon intrinsic signalling ($I$), while manipulation of stromal cells or tumour-stroma signalling (such as hormonal therapy for localised disease or a bisphosphonate for bony metastatic disease) would preferentially effect the $D$ cells.  Additionally, as the steady states depend not only on the parameters (which can be manipulated by therapy) but also on the relative populations, the timing of therapy can drastically change the results.

\begin{figure}[!htb]
\centerline{\includegraphics[width=9cm]{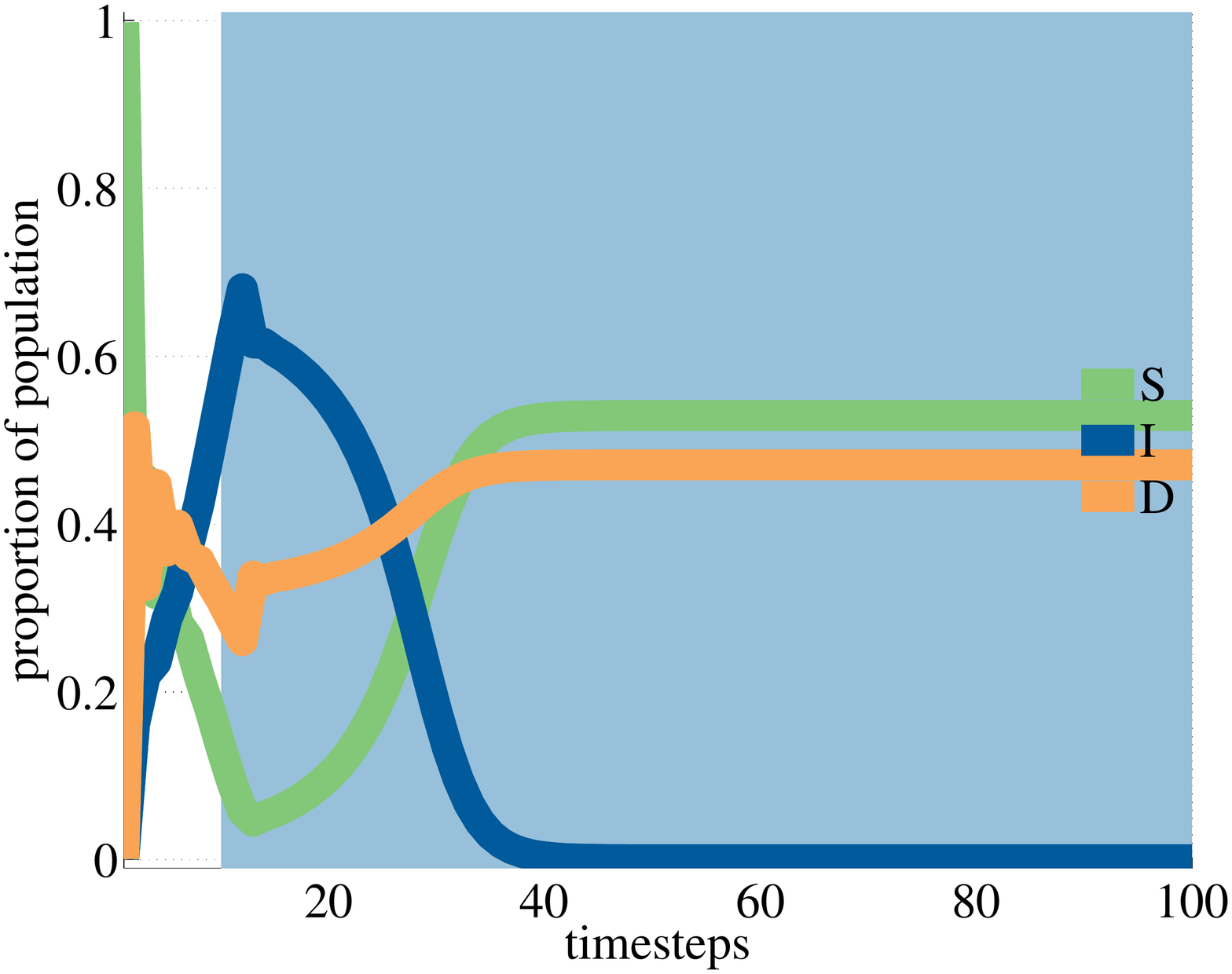}\includegraphics[width=9cm]{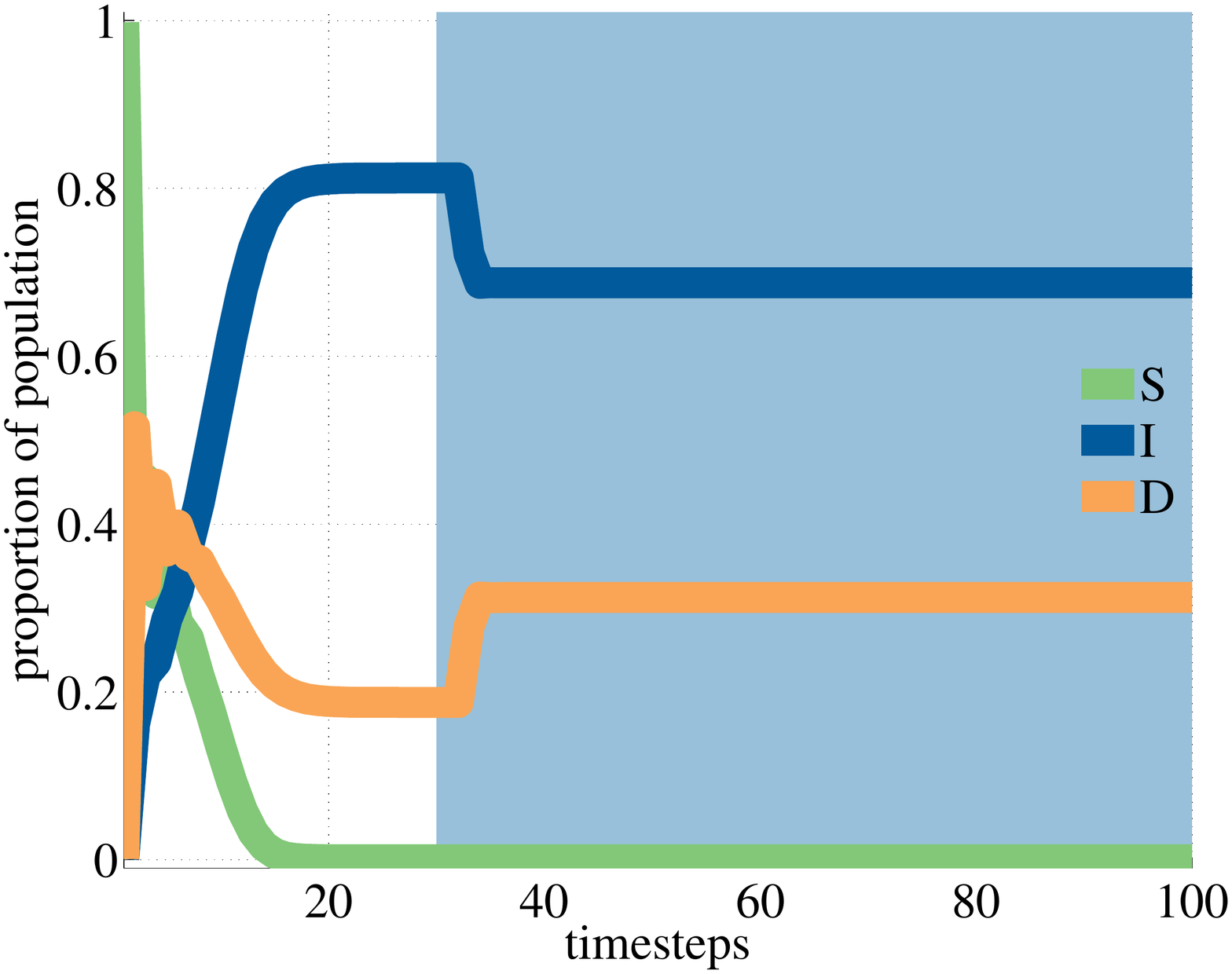}}
\caption{\label{fig:biothe}The importance of timing. In all cases we start the simulations with $\alpha=0.5, \beta=0.7, \rho=0.1$ and $\gamma=0.8$. \textbf{Left} we show an early initiation with a therapy directed at stromal independent cells which effectively destroys the population before it can take over. \textbf{Right} we show the same therapeutic intervention but initiated at a later time and there is an effect, but it does not change the biology of the tumor fundamentally.}
\end{figure}

An exploration of a putative biological therapeutic agent with different timing strategies is represented in figure \ref{fig:biothe}.  In the case of this simulated biological therapy (possibily representing an mTOR inhibitor - which preferentially penalises the $I$ population), the end effect of the therapy is strongly influenced by the time of initiation - when initiated early (when there is still competition between $I$ and $D$), it drastically alters the outcome of the game, while late application (after $I$ has already dominated) only causes a small shift.  This result can be seen as an application of the \emph{kairos principle} and speaks to the importance of choosing the right therapy at the right time.  Testing this same concept in a more 'stromal targetted' therapy (e.g. hormonal manipulation) did not produce significant differences, suggesting that a short delay in onset is of less importance in this therapeutic strategy.  This result does not suggest that early versus late hormone therapy is meaningless - in fact it has been shown that early hormone therapy can slightly increase overall survival - but instead that the timing will not effect the overall outcome of the game.

Testing a different therapeutic strategy, the duration of 'stromal directed' therapy is shown in figure \ref{fig:hormone}.  Here, we hypothesize that application of such a therapy would primarily effect the benefit $D$ cells receive from the interatcion with $S$ cells (i.e. $\alpha$ will be reduced).  This results in a shift in the opposite direction from the prior therapy - from $D$ to $I$.  Further, we see that the duration of therapy strongly effects the end result of the game.  The final result (figure \ref{fig:hormone}c) recapitulates the clinical reality of evolution of resistance to hormonal therapy (e.g. castrate resistant prostate cancer).  

\section{Discussion}

Although many nascent prostate tumours never become life threatening, those that do can use two different and distinct routes: either becoming microenvironmentally independent (representing a small cell prostate cancer type scenario) or by co-opting certain stromal cells in order to sustain tumour progression (as happens in bony metastatic disease). We present a simple model that abstracts key aspects of prostate cancer evolutionary dynamics including progression towards lethal outcomes that can be either stromagenic (resulting from mutualistic interactions between the tumour and certain stromal cells) or stromal independent. These outcomes, reflecting evolutionary changes of the genetically unstable, heterogenous tumor cell population, are influenced by the interactions both between the different populations ($I$, $S$ and $D$) and with their microenvironment. Primary tumours are likely to contain areas that are stromagenic as well as other areas that are stromal independent which would make application of the therapeutical message of this model less relevant. On the other hand secondary sites are thought to represent clonal populations originating from a specific (either stromagenic or stromal independent) area of the primary tumour \cite{Navin:2011fk}, therefore extrapolating to the metastatic situation is more appropriate.

We have previously demonstrated \cite{Anderson:2009eu} dominance of $D$ phenotypes happens naturally in a microenvironment rich in resources whereas resource-poor microenvironments tend to select for $I$ cells. These results are further validated by this model: regardless of the absolute fitness of the $I$ phenotype (as given by $\gamma$), an increase in the proportion of $I$ cells is observed as the microenvironment becomes resource-poor (as given by $\beta$, right column of figure \ref{fig:outcomes}). Stroma and its interactions with the tumour have a tremendous impact on the phenotypic composition of the tumour. In those cases in which $\alpha$ is sufficiently high, denoting stroma which can be supported by the tumour, the cooperation between $D$ and $S$ can push the $I$ population towards extinction (see figure \ref{fig:example} left panel). Conversely, this can be described as tumor cells which are particularly effective at eliciting stomal support - the key is the interactive process, not so much the cell's absolute behavior.  

On the other hand, when $\rho$ is sufficiently high, the ensuing mutualistic relationship between $D$ and $I$ can lead to dominance of the tumour phenotypes at the expense of $S$, even when the $D$ cells can derive some benefit from their interactions with the stroma (see figure \ref{fig:example} right panel). The least frequent of the outcomes, coexistence of the three phenotypes, depends on values of $\alpha$ being high enough to promote the collaboration between $D$ and stromal cells but not so high that $I$ cells are driven to extinction. Higher values of $\rho$ promote polyclonal tumours as they allow $D$ cells to coexist with $I$ cells. Thus, those tumours in which the $D$ cells can benefit from the factors produced by $I$ cells and still cooperate with the stroma are more likely to sustain the three phenotypes in an evolutionarily stable strategy.

\begin{figure}[!htb]
   \centerline{\includegraphics[height =6cm,width=6cm]{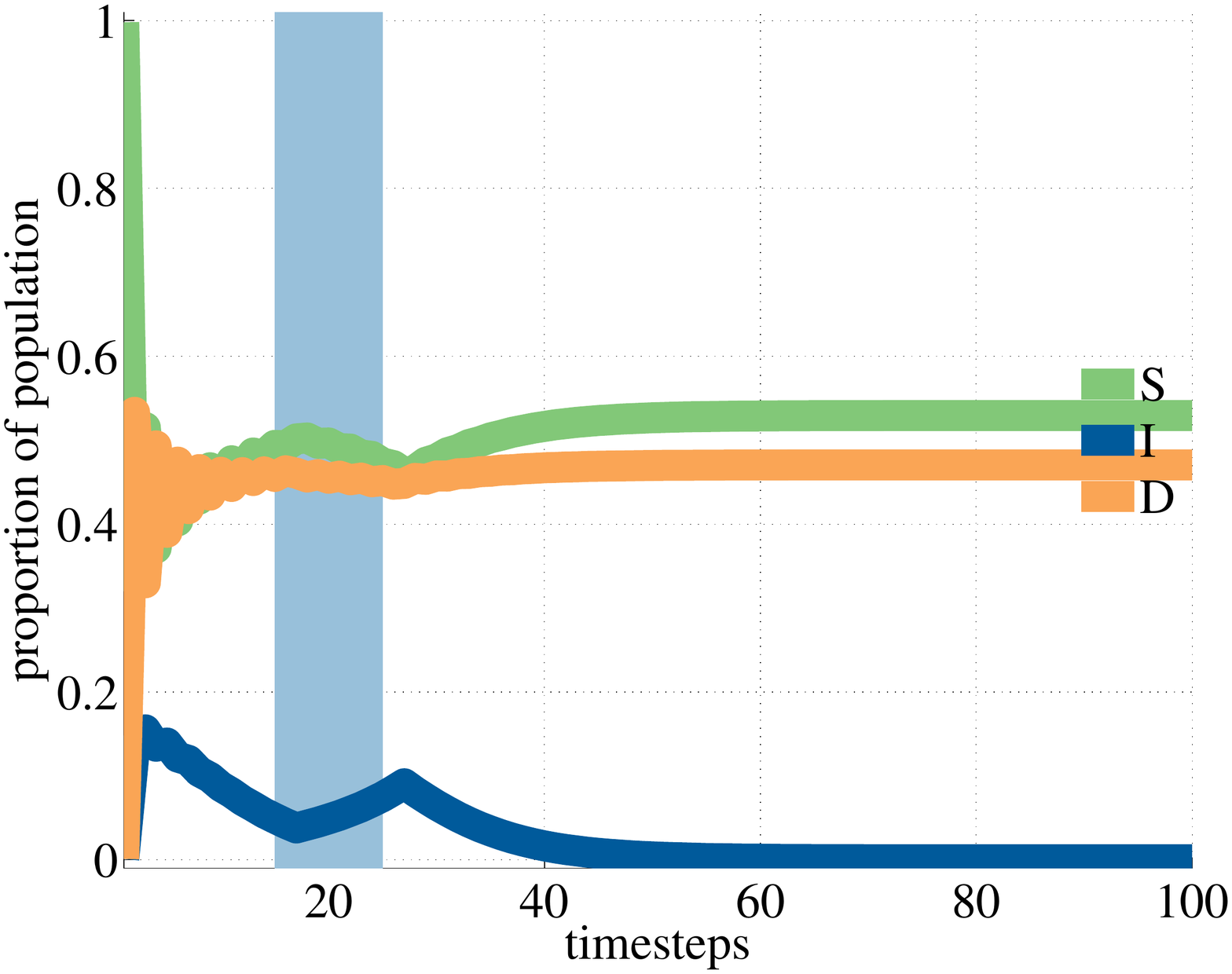}\includegraphics[height =6cm,width=6cm]{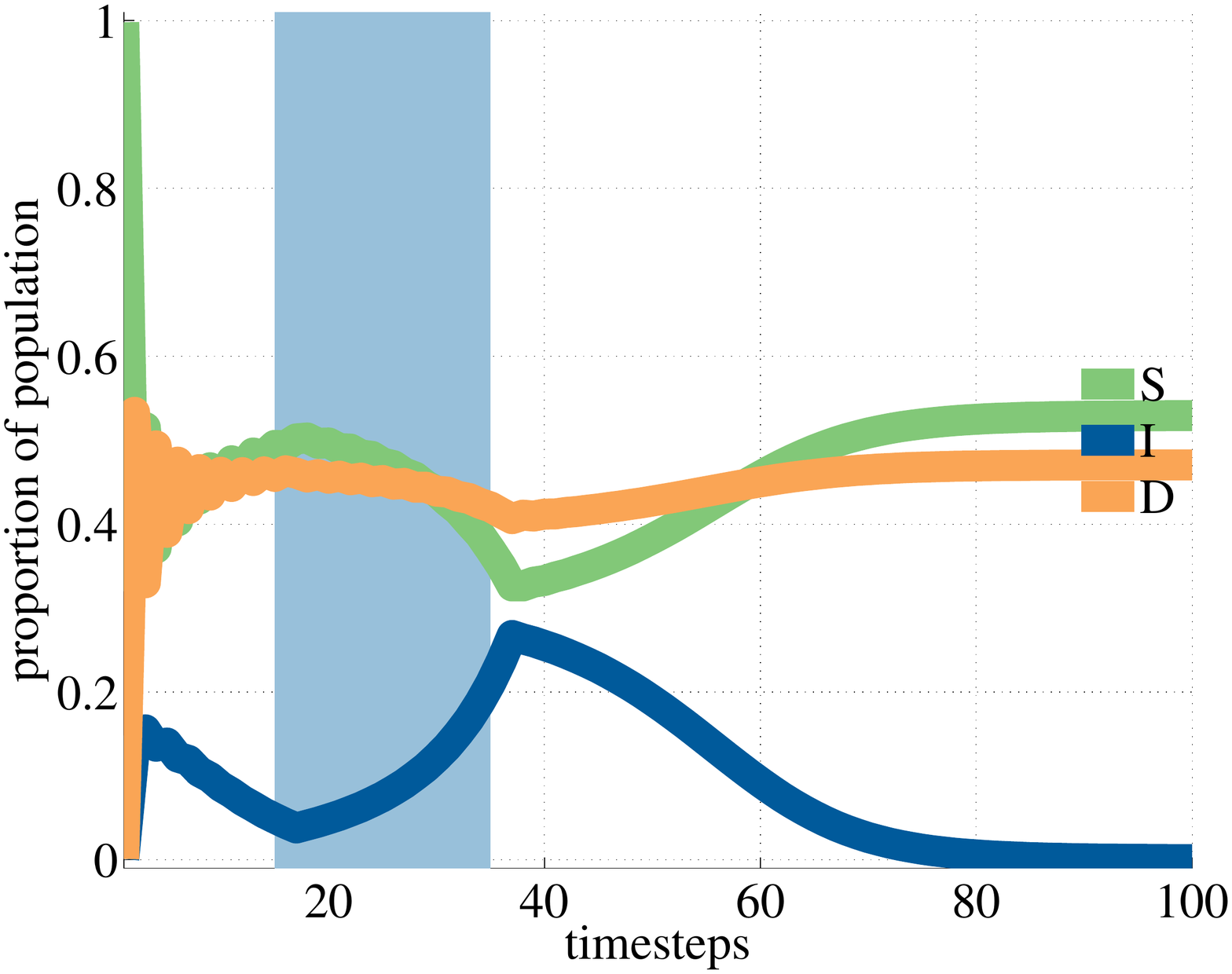}\includegraphics[height =6cm,width=6cm]{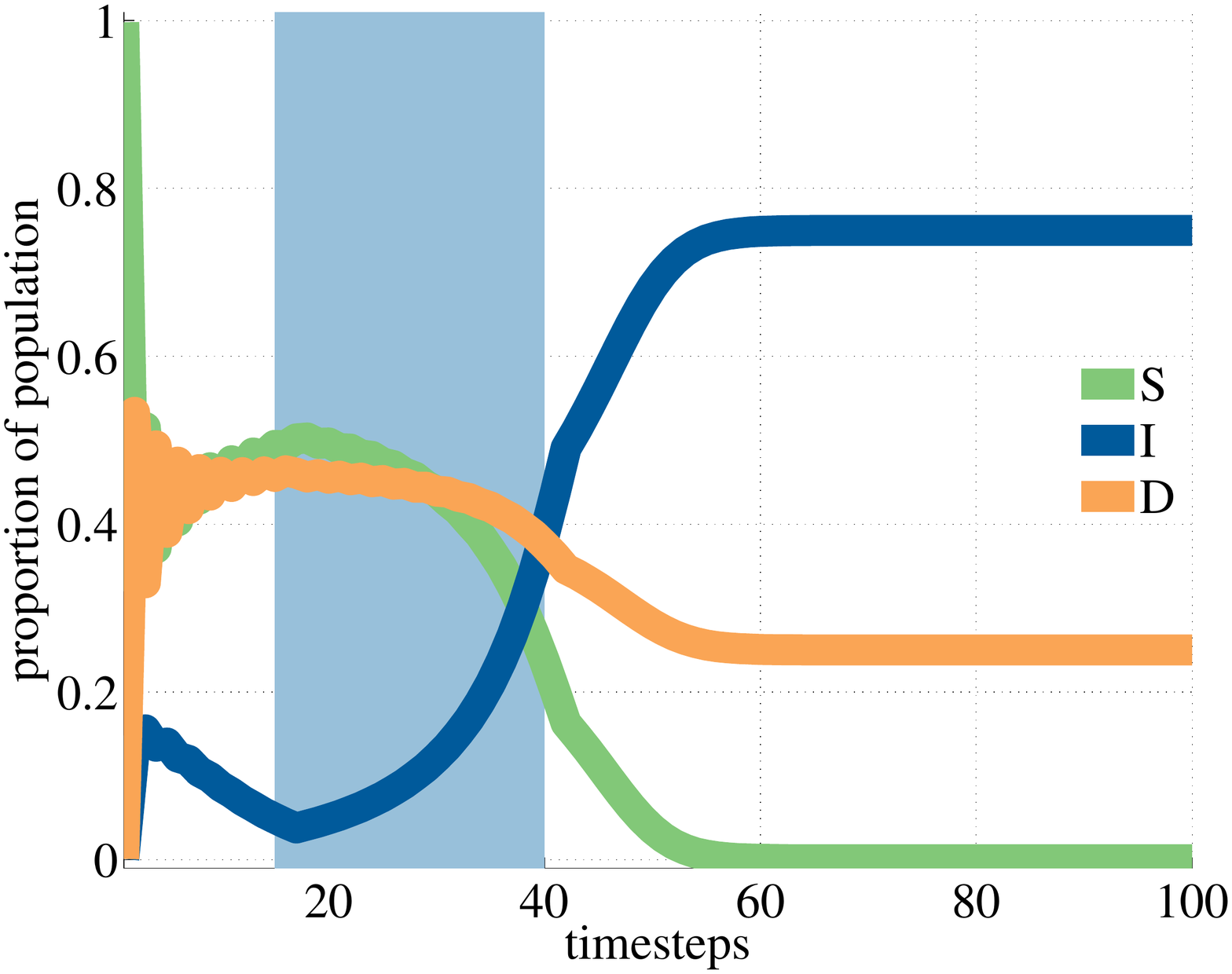}}
\caption{\label{fig:hormone}Hormone therapy. In all cases we start the simulations with $\alpha=0.5, \beta=0.7, \rho=0.1$ and $\gamma=0.8$. \textbf{Left} initiation at timestep 15 with a short duration pulse  of 10 timesteps of 'stromal directed' therapy shows a very minimal effect and no change in the steady state.  \textbf{Center} initiation at timestep 15 with a medium duration pulse of 20 timesteps of 'stromal directed' therapy shows a substantial effect that slowly returns to the initial steady state.  \textbf{Right} initiation at timestep 15 with a long duration pulse of 25 timesteps of 'stromal directed' therapy shows a substantial effect that changes the overall steady state and effectively changes the biology of the tumor - recapitulating the emergence of resistance to therapy.}
\end{figure}

To combat the emergence of castrate resistance (and side effects) in the treatment of metastatic prostate cancer, many physicians have adopted non-standard dosing schedules for their androgen ablation therapies.  A recent study suggested that either using the patient's measured testosterone levels as a guide, or using an intermittent schedule as opposed to constant dosing (a calendar schedule) slowed the onset of castrate resistance.\cite{blumberg:2011}  While this study has shown that alternative schedules can provide benefit, there is, as of yet, no standard of care for this.  To explore this question, simulations were run with differing times of initiation and duration of 'stromal directed' therapy.  The results of the game given different durations of 'stromal directed' therapy show significant, fundamental differences in outcome with differing schedules.  Initiation with short duration of treatment (figure \ref{fig:hormone}a) results in no appreciable change in the game.  Initiation for long duration (figure \ref{fig:hormone}c), as expected, shows the evolution of castrate resistance and a phase transition to a new steady state dominated by I cells which will then be insensitive to further manipulation with this therapy, requiring a strategy change.  There is, however, an optimal duration (figure \ref{fig:hormone}b) where the $D$ cells are reduced without an irreversible increase in the $I$ cells.  When the therapy is taken off, the levels begin to return to the original $D$ dominated steady state; a situation where 'stromal directed' therapy will work again.  

These results, taken together, suggest that different therapies are likely of different value to the two major tumour types:  $D$ tumours are likely best treated with stromal manipulation (e.g. hormonal therapy) while $I$ tumours would be better treated initially with a biologic agent such as an mTOR inhibitor.

Regardless of timing and schedule, the emergence of castrate resistance is largely unavoidable given a long enough time frame, and these results suggest that when this emergency does occur, a treatment strategy change is in order.  These sentiments are echoed in the recent literature which suggests that the addition of mTOR inhibitors to hormone therapy after the onset of castrate resistance\cite{Schayowitz:2010fk} could hold promise. 

\section{Conclusion}

The results presented here highlight the importance of understanding tumour-stroma interactions in driving not only tumour outcomes (whether malignant or not) but also their impact on potentially new therapeutic approaches. A key assumption made here is that the biology of metastatic sites is related to the original biology of the primary tumor.  While this assumption carries with it many other assumptions, it is one that underlies the majority of clinical decision making and biological extrapolation in prostate cancer.  

In the clinical paradigm of metastatic prostate cancer, which has recently been muddied with many new agents coming to the arena, it has become ever more important to develop rational strategies with which to test new sequence and timing regimens.  For example, it may be that the time honored standard of hormonal manipulation until failure is no longer the best strategy - but how can we rationally integrate these new therapies?  While our results provide a theoretical framework with which to understand clinical observations they do not yet provide us with the ability to make strong recommendations for timing and duration of therapy (limited by the arbitrary nature of the time variable in this model), however, this model has already generated testable hypotheses and with direct experimental parameterization and validation could lead to specific therapeutic strategies.

\section{Acknowledgements}
We gratefully acknowledge the NIH/National Cancer Institute support from the ICBP (5U54 CA113007), PSOC (U54 CA143970-01) and TMEN (1U54 CA126505) programs.
\bibliography{mmbMiamiJGS.bib}
\bibliographystyle{unsrt}

\end{document}